\def\aa{{A\&A}}
\def\aas{{A\&AS}}
\def\aj{{AJ}}

\def\apj{{ApJ}}
\def\apjs{{ApJS}}

\def\mnras{{MNRAS}}

\documentclass[cup5b]{caps}
\usepackage{epsfig}
\usepackage{graphicx}
\usepackage{amssymb}
\usepackage{ociwsymp1e}
\HeadText{F. Panessa et al.}

\begin{document}

\pagenumbering{arabic}

\author[]{F. PANESSA$^{1,2}$, M. CAPPI$^{1}$, L. BASSANI$^{1}$, M. 
DADINA$^{1}$, R. DELLA CECA$^{3}$, \cr S. PELLEGRINI$^{2}$, 
G. TRINCHIERI$^{3}$, A. WOLTER$^{3}$, and G. G. C. PALUMBO$^{2}$
\\
(1) IASF-CNR, sezione di Bologna\\ 
(2) Universita' di Bologna\\
(3) Osservatorio Astronomico di Brera, Milano}

\chapter{The Nature of ``Composite" Seyfert/Star-Forming Galaxies}

\vspace{1cm}

\begin{abstract}

We present the results obtained with {\it Beppo}SAX observations of the three 
Composite Seyfert/star-forming galaxies: IRAS 20051-1117, IRAS 04392-0123 
and IRAS 01072+4954. These sources belong to an 
enigmatic class of X-ray sources detected in the ROSAT All-Sky Survey
(Moran et al. 1996) which is composed of 6 low redshift galaxies.
Their optical spectra are dominated by the features of H~II galaxies while 
their X-ray luminosities ($\geq$ 10$^{42}$ ergs/s) are typical of Seyfert galaxies.
IRAS 20051-1117 shows a 2-10 keV spectrum well
described by a power-law with $\Gamma$ = 1.9 and low intrinsic absorption. 
This result, the ratio Flux(2-10 keV)/Flux([O~III]$\lambda$5007) and the significant
X-ray variability detected, clearly rule out a Compton
thick nature of this source. IRAS 04392-0123 and IRAS 01072+4954, instead, 
have a {\it Beppo}SAX flux a factor of $\sim$
50-80 smaller than in the previous ROSAT observations, resulting in poor statistics,
that prevents detailed modeling.
 
\end{abstract}

\section{Introduction}

A large spectroscopic optical survey of bright IRAS and X-ray selected sources from the ROSAT 
All Sky Survey revealed an enigmatic class of 6\footnote{The original Moran et al.'s list
was composed of 7 objects, but IRAS 10113+1736 is no longer a valid candidate, since infrared
and X-ray emission originate from different sources (Condon et al. 1998).} 
low redshift galaxies with optical spectra
dominated by the features of H~II galaxies but X-ray luminosities typical of AGNs,
ranging from 1.5 $\times$ 10$^{42}$ erg/s to 5 $\times$ 10$^{43}$ erg/s 
in the ROSAT band (Moran et al. 1996; see Table~\ref{sample}). 
These galaxies were named ``Composite''. 
The diagnostic emission line ratio diagrams (Veilleux \& Osterbrock 1987) classify 
these objects as star-forming galaxies. Yet, some of them present [O~III]$\lambda$5007 
lines significantly broader than all other narrow lines in the spectrum and weak 
and elusive broad H$\alpha$ wings. Both evidences suggest the presence of a weak 
or ``obscured'' AGN. An X-ray spectrum is available only for IRAS 00317-2142 
observed by ASCA (Georgantopoulos 2000). Its X-ray spectrum
can be well reproduced by a single power-law with $\Gamma$ = 1.7, with low
absorption and no detected iron line at 6.4 keV.  

Other similar galaxies (i.e. with bright X-ray emission but
weak or absent AGN features in the optical band) have been found also in deep ROSAT
fields (Boyle et al. 1995, Griffiths et al. 1996) and in the {\it Chandra} 
and XMM-{\it Newton} deep-fields (Rosati et al. 2001, Fiore et al. 2000, 
Severgnini et al. 2003).  Although these 
sources usually lack evidence for strong starburst emission, the main problem
is the same as for the Moran et al. (1996) sources, i.e., to be able
to explain the optical weakness/disappearance of the AGN in these X-ray 
luminous sources.

It is not likely that the starburst could overpower a Seyfert optical nuclear spectrum, 
since the starburst component in these objects is not particularly strong
(Moran et al. 1996). To explain the absence of optical Seyfert lines we consider
three different scenarios: I) the nucleus is heavily obscured and also the NLR is obscured,
II) the absorber surrounds almost completely the AGN preventing the UV flux from
ionizing the matter otherwise responsible for optical lines and diminish the strength of the
Seyfert emission lines, III) The BLRs and the NLRs are absent (Panessa \& Bassani 2002).
Furthermore it is worth note that HST observations of a galaxy previously classified as 
starburst/H~II on 
the basis of ground-based observations, have recently revealed a low luminosity 
Seyfert 2 nucleus hidden by a strong nebular emission from H~II regions 
near the nucleus (Gezari et al. 2003).

 \begin{table}
    \caption{Moran et al. (1996) Composite sample}
    \begin{tabular}{lcccc}
     \hline \hline
     {Name}  & {$N_{HGal}$} & {$F_{0.1-2.4 keV}$} & {$F_{[O~III]}$} & {$F_{IR}$} \\
   & {$10^{20} cm^{-2}$} & {$10^{-12} (cgs)$} & {$10^{-14} (cgs)$} &  {$10^{-10}
(cgs)$} \\
     \hline
     IRAS 00317-2142 & 1.47 & 3.88 & 2.26 & 2.31\\
     IRAS 00374+5929 & 7.84 & 2.95 & 0.85 & 0.61 \\
     IRAS 01072+4954 & 13.95 & 1.64 & 6.39 & 0.50\\
     IRAS 01319-1604 & 1.39 & 1.00 & 1.75 & 1.13\\
     IRAS 04392-0123 & 5.62 & 2.13 & 1.83 & 0.49 \\
     IRAS 20051-1117 & 6.53 & 1.30 & 1.27 & 1.20 \\
        \hline \hline
    \end{tabular}
  \label{sample}
 \end{table}

\section{Results}

We have obtained {\it Beppo}SAX observations for three sources.
Their fluxes and luminosities are listed in table~\ref{flux}.
In the following sections we discuss each source.

  \begin{figure}
    \centering
    \centerline{\epsfig{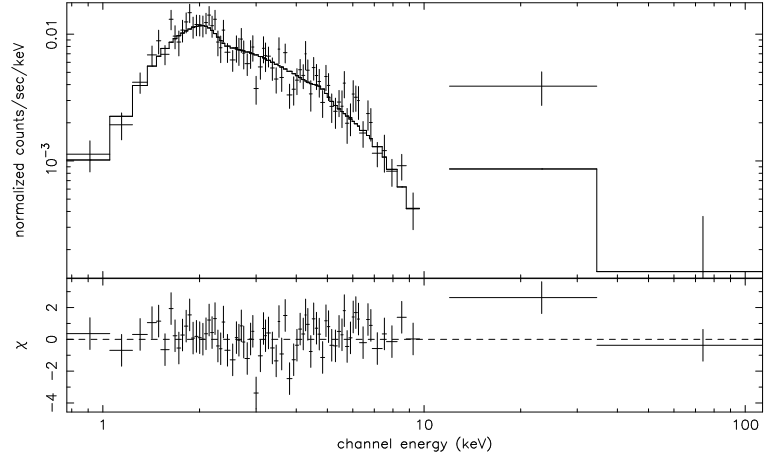}}
    \caption{MECS23 1-10 keV spectrum of IRAS20051-1117}
    \label{spec}
  \end{figure}

\subsection{IRAS 20051-1117}

The 2-10 keV spectrum of this source is well described by a single power-law model with 
$\Gamma$ = 1.9 $\pm{0.1}$ and intrinsic absorption $\leq$ 1.2 $\times$ 10$^{21}$ cm$^{-2}$ 
in excess to the Galactic value. An iron K$\alpha$ line at 6.4 keV is not 
statistically significant; we measure an upper limit for the equivalent width of 226 eV (90\%). 
The 4-10 keV lightcurve shows a significant variability 
(a factor of $\sim$ 3 in a time scale of $\sim$ 3.6 ks).  

The X-ray, infrared and optical properties of this object are similar to NGC 7679
(Della Ceca et al. 2001) and IRAS 00317-2142 (Georgantopoulos 2000): unabsorbed X-ray emission
but possible absorbed (or intrinsecally weak) optical emission from an AGN.
It is likely that in these objects the central Seyfert nucleus produces the strong X-ray emission 
while the surrounding star-forming region produces the strong Far-Infrared emission, but it
is not clear why we cannot see the Seyfert emission lines. 
The surrounding starburst could provide sufficient obscuration to cover the optical emission 
regions (Levenson et al. 2001). It is also possible that the starburst and/or the absorber 
completely surrounds the nucleus with almost spherical geometry thus preventing 
the UV nuclear flux from ionizing the matter. However, 
the ratio Flux(2-10 keV)/Flux([O~III]$\lambda$5007) = 177.2, the lack of a strong iron line and 
the steep slope of the power-law all suggest a Compton thin nature for IRAS 20051-1117, supporting
the scenario of the lack of NL and BL regions.

\subsection{IRAS 04392-0123}

The spectrum of this object has very poor statistics 
($\leq$ 116 counts for the MECS). The 2-10 keV spectrum is well 
described by a single power-law with $\Gamma$ = 1.7 $\pm{0.6}$. The absorption is 
here only poorly constrained with an upper limit of 7.6 $\times$ 10$^{22}$ cm$^{-2}$. 
The 0.1-2.4 keV flux is lower by a factor of
$\sim$ 50 with respect to the ROSAT flux. 
The Flux(2-10 keV)/Flux([O~III]$\lambda$5007) ratio is 10.4.

\begin{figure}
    \centering
    \centerline{\epsfig{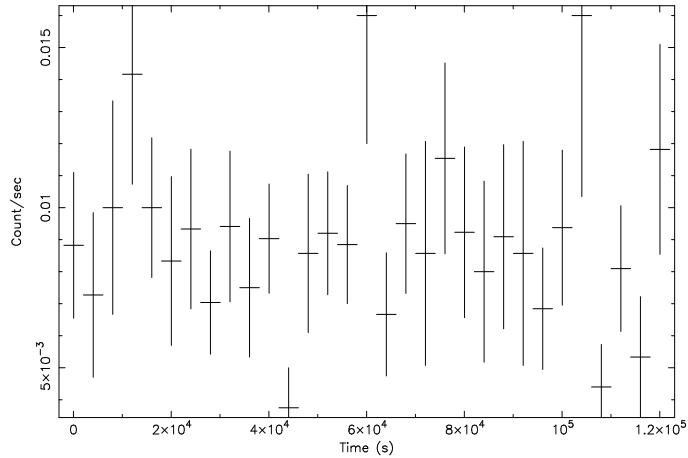}}
    \caption{MECS23 4-10 keV lightcurve of IRAS20051-1117}
    \label{lc}
  \end{figure}

 \begin{table}
  \caption{BeppoSAX fluxes and luminosities}
    \begin{tabular}{lcccc}
     \hline \hline
     {Name}  & {$F_{0.1-2.4 keV}$} & {$F_{2-10 keV}$} & {$F_{10-50 keV}$} &{$logL_{2-10 keV}$}   \\
   & {$10^{-12} (cgs)$} & {$10^{-12} (cgs)$} & {$10^{-12} (cgs)$} &  \\  
     \hline
     IRAS 20051-1117 & 1.68 & 2.25 & 3.6 & 42.57 \\
     IRAS 04392-0123 & 0.03 & 0.19 & - & 41.46 \\
     IRAS 01072+4954 & 0.02 & 0.17 & - & 41.24 \\
     \hline \hline
    \end{tabular}
  \label{flux}
 \end{table}

\subsection{IRAS 01072+4954}

Due to the small number of counts ($\leq$ 40 counts for the MECS) for 
this source we assume a power law model
with Galactic absorption to estimate the 
0.1-2.4 keV and 2-10 keV fluxes. The 0.1-2.4 keV flux is lower by a factor of 
$\sim$ 80 compared to the ROSAT flux. The ratio Flux(2-10 keV)/Flux([O~III]$\lambda$5007) 
is 2.7.

\section{Future work}

{\it Chandra} observations of 4 composites of the Moran et al. (1996) sample have been
awarded to our group (Cycle 4, PI: A. Wolter). Thanks to the {\it Chandra} high resolution
it will be possible to resolve the nuclear X-ray emission and investigate the presence
of a diffuse soft thermal component which could be associated with the starburst.

\end{document}